# Proof of operation of a micro-hollow cathode discharge driven by a true ns–pulsed high voltage: emission properties and atomic nitrogen production

N. Chazapis[1], D. Stefas[1], A. Brisset[2], L. Invernizzi[1], N. Girodon-Boulandet, G. Lombardi[1], C. Lazzaroni[1], K. Gazeli[1,a)]

[1]*Laboratoire des Sciences des Procédés et des Matériaux (LSPM), CNRS, Université Sorbonne Paris Nord, UPR 3407, F–93430 Villetaneuse, France*
[2]*Laboratoire Energétique Moléculaire et Macroscopique, Combustion (EM2C), CNRS, Université Paris Saclay, UPR 288, F–91190 Gif-sur-Yvette, France*

[a)]Author to whom correspondence should be addressed: kristaq.gazeli@lspm.cnrs.fr

## Abstract

The implementation of Micro-Hollow Cathode Discharges (MHCDs) for the efficient and controlled dissociation of molecular nitrogen into reactive N–atoms remains a central challenge in the synthesis of strategic materials such as hexagonal boron nitride (h-BN). Traditional MHCDs driven with DC high voltages (HV) can generate N–atoms under low-to-moderate gas pressures. However, they are prone to intrinsic instabilities that eventually induce arcing, increase thermal load and limit significantly their lifetimes. In this work we aim to enhance $N_2$ dissociation efficiency in an Ar/$N_2$ MHCD driven by a true nanosecond (ns) pulsed positive HV which has not been yet considered in MHCDs. A testbed reactor is employed to investigate microplasma behavior under ns-pulsed operation, which is then compared to a conventional DC MHCD. Specifically, we combined optical emission and ns-TALIF diagnostics under suitable operating conditions of each HV regime comprising different gas mixtures, pressures, and electrical parameters. These measurements allowed for the identification of key excited species and mapping of ground-state N–atoms absolute density in the low-pressure chamber of the MHCD reactor. It is demonstrated that ns-pulsed excitation produces a more symmetric discharge expansion on the cathodic surface compared to a standard DC excitation, while generating similar nature of emissive species and consuming up to 2 orders of magnitude lower average power depending on the operating frequency and voltage. Importantly, up to 6–fold enhancement in absolute N–atoms density (maximum value measured: $1.14\times10^{15}$ cm$^{-3}$) is achieved with the ns-pulsed MHCD making it very promising for h-BN synthesis.

Micro-hollow cathode discharges (MHCDs), first introduced 30 years ago by Schoenbach *et al.*,[1] are microplasma sources with fundamental interest and emerging applications in material synthesis, photonics, and microelectronics.[2–4] A distinct type of MHCD consists of two electrodes separated by a dielectric layer and a central hole of several hundred micrometers in diameter.[5–9] Positioning this structure at the interface between two chambers which are maintained at the same or different pressure provides a favorable configuration for generating a large-volume plasma in the low-pressure chamber.[2,3,8,9] This device consumes very low electric power (only few W or lower) but effectively generates huge power densities inside the hole (up to 100 kW cm$^{-3}$) due to its micrometric dimensions. This enables enhanced electron production (up to $10^{15}$ cm$^{-3}$)[5,10,11] which triggers a rich plasma chemistry via excitation, ionization and dissociation of key reactive species. These characteristics are particularly attractive for nitride material synthesis, such as hexagonal boron nitride (h-BN), which requires the efficient dissociation of highly stable molecules such as $N_2$.[3]

$N_2$ is an abundant precursor of reactive N–atoms which has been widely adopted in MHCDs for h-BN synthesis.[2,3] However, its strong triple bond is an important bottleneck for its efficient dissociation. In order to enhance N–atom production, a comprehensive understanding of the effect of gas composition and electrical excitation parameters (voltage type and amplitude, frequency, etc.) on the MHCD physics/chemistry is required. The most common and widely studied means to generate MHCDs is via a DC high voltage (HV) excitation, which already has been demonstrated to generate N–atom densities as high as $10^{15}$ $cm^{-3}$ depending on the system parameters.[5,12] However, traditional DC MHCDs are susceptible to arcing and exhibit limited operational lifetimes as well as intrinsic instabilities, including often uncontrolled transitions between abnormal, normal glow, and self-pulsing discharge regimes.[10,11,13] These drawbacks largely stem from the continuous exposure of device materials to reactive plasma neutrals, ions and elevated thermal loads. These can be partially mitigated by driving MHCDs with repetitive sub-µs HV pulses,[14–16] which have been found effective in synthesizing h-BN on 2–inch Si substrates.[2,3,17] However, absolute N–atom densities have not yet been measured in pulsed-driven MHCDs; their presence has far been inferred only indirectly via OES diagnostics.[3,16] Thus, quantifying and controlling the production of N–atoms in pulsed MHCDs would be a great asset for their optimization for h-BN synthesis.

In this work, we extend beyond conventional DC and sub-µs pulsed MHCDs by investigating the electrical and optical characteristics of an Ar/$N_2$ MHCD driven by true nanosecond (ns) HV pulses. Experiments are performed in a dedicated MHCD reactor (**Figure 1a**) across a broad range of Ar/$N_2$ gas mixtures, pressures and voltage amplitudes. Its performance is compared with that of a DC MHCD to assess its potential for enhancing N–atom production while reducing thermal loading and improving source longevity.

The MHCD operates both in continuous DC and ns-pulsed HV excitation. It consists of two circular molybdenum electrodes (20 mm diameter, 100 µm thickness) separated by a circular alumina ($Al_2O_3$) dielectric disk (35 mm diameter, 750 µm thickness), as shown in **Figure 2(a)**. These are glued together, and a central hole (approximately 300 µm in diameter) is drilled through the assembly. The MHCD is positioned inside a reactor at the interface between two chambers maintained at equal or different pressures and connected through the central hole (**Fig. 1(a)**). In the case of using different pressures, an Ar/$N_2$ gas mixture is introduced into the high-pressure ($P_{High}$) chamber (≥30 mbar), while the low-pressure ($P_{Low}$) chamber (≤10 mbar) is connected directly to a vacuum pump. For the DC operation, the electrode facing the low-pressure chamber serves as the anode and is grounded, whereas the electrode facing the high-pressure chamber serves as the cathode and is connected to the output of a negative HV power supply (T1070006, THQ, ISEG). To regulate the discharge current and suppress arcing, a 480 kΩ ballast resistor ($R_b$) is integrated between the high-voltage source and the cathode. The net discharge voltage ($V_{dis}$) is monitored via a voltage probe positioned downstream of $R_b$; $V_{dis}$ is smaller than the incident voltage ($V_{inc}$) which is set on the negative power supply.[6] Furthermore, a 180 Ω resistor ($R_m$) is connected between the anode and the ground to measure the discharge current ($I_{dis}$) via a second voltage probe connected on the terminals of $R_m$. Because $V_{dis}$ and $I_{dis}$ are strictly governed by the plasma features, they cannot be varied independently.[6]

Under ns-pulsed operation, the output of a HV power supply (FID, FPG 30-30NM4), delivering positive HV pulses with variable features (duration<10 ns, rise time<2 ns, repetition frequency $f$=0.001–10 kHz, voltage amplitude $V_P$=1–20 kV), is connected to the electrode located on the low-pressure side (anode), while the opposite electrode, found on the high-pressure side (cathode), is grounded. In this case, no ballast resistor is needed, since the average current is small enough to limit arcing, which is already an important advantage over DC operation. Due to impedance mismatch between the coaxial cable delivering the voltage and the MHCD, electrical signals are reflected on the electrodes and generator, and secondary pulses are produced.[18] Before breakdown, the MHCD exhibits a very high impedance. This drops dramatically after plasma ignition, representing a variable low-impedance resistive load. The electrical waveforms are recorded using a LeCroy HV probe (PHV4–4051) and a current transformer (Pearson 6585), positioned 3 m away from the discharge. $V_{dis}$ and $I_{dis}$ signals are reconstructed from the incident and reflected pulses measured at the same position, as described in.[18,19] For both HV excitations, $V_{dis}$ and $I_{dis}$ serve for the calculation of the consumed average power and corresponding power density.

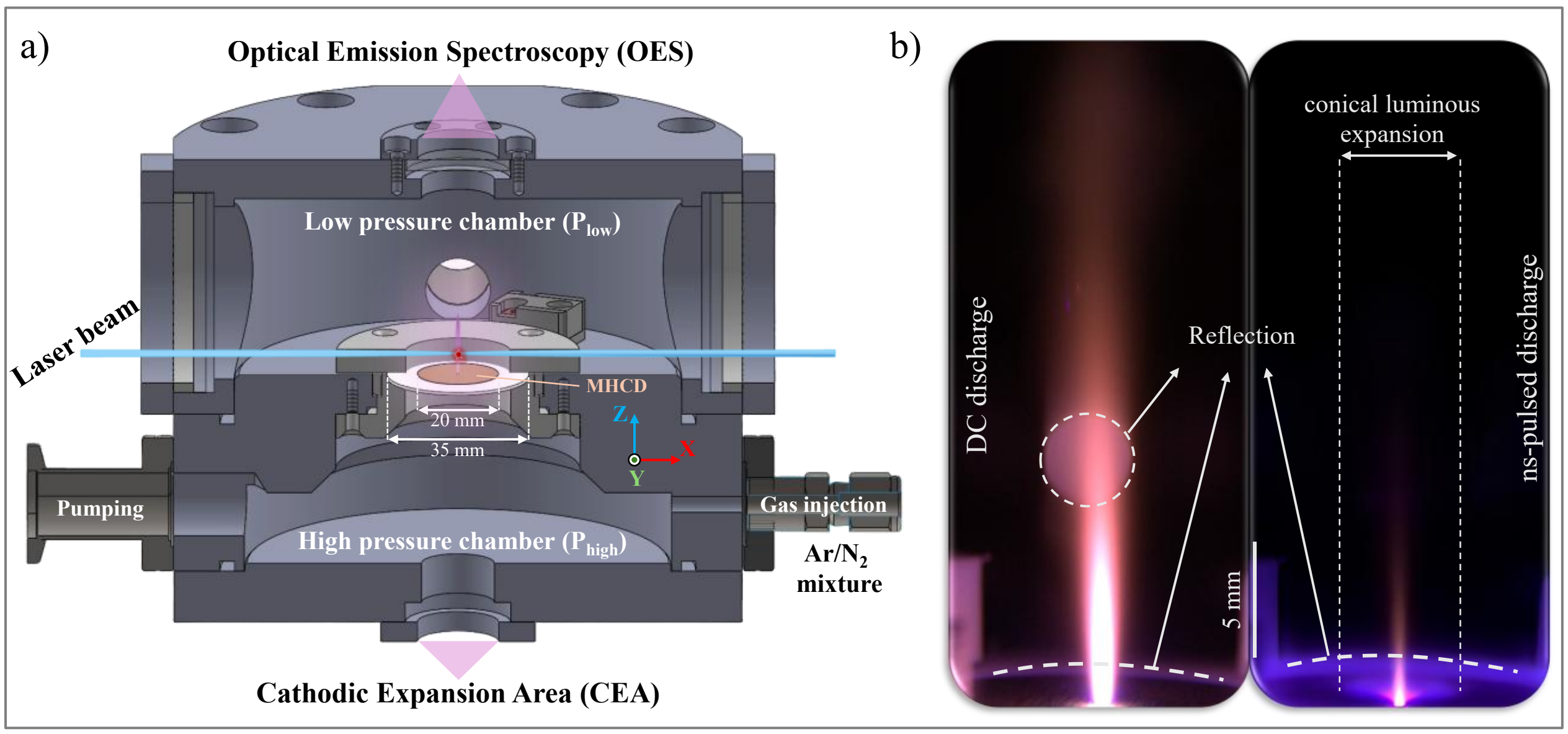

**FIG. 1.** (a) MHCD reactor and optical diagnostics used. (b) Representative DSLR photos (1 s exposure time) of the discharge expansion in the low-pressure chamber under DC ($V_{dis}$=360 V, $I_{dis}$=2 mA) and ns-pulsed ($V_{dis}$=3.7 kV, $f$=1 kHz) excitations ($P_{Low}$=13 mbar, $P_{High}$=47 mbar). The dotted circle and arcs indicate reflections of the effluent on the rear window of the chamber, and the discharge generated on the cathode side, respectively.

When a pressure differential is imposed between the two chambers and electrodes are biased by an HV, a microplasma ignites within the MHCD hole and expands into the low-pressure chamber as a distinct luminous effluent (**Fig. 1(b)**). When maintaining an equal gas pressure between the two chambers, the microplasma is confined within the hole. In both cases, a discharge also expands on the cathode surface.[7] In this study both conditions are considered by characterizing corresponding microdischarges through three methods (**Fig. 1(a)**): *(i)* space-time integrated OES to identify the dominant excited species; *(ii)* conventional Digital Single Lens Reflex (DSLR) photography for assessing the discharge cathodic expansion area (CEA); and *(iii)* nanosecond two-photon absorption laser-induced fluorescence (ns-TALIF) to quantify the spatio-temporal evolution of ground-state N–atoms in the low-pressure chamber. **Fig. 1(b)** shows representative DSLR photos of the luminous effluent in the low-pressure chamber under both HV waveforms. DC discharge generates a longer, thicker and brighter effluent due to the continuous transport of excited species downstream exclusively by gas flow.[12] However, in the ns-pulsed case the effluent is shorter, narrower and less bright. This difference should be partially attributed to the exposure of the gas mixture to a significantly lower number of discharges in the ns-pulsed case. For instance, its corresponding pattern in **Fig. 1(b)** (1 s integration time) originates from only $10^3$ HV pulses (1 kHz frequency). Assuming that each HV pulse starts immediately after the previous one, which mimics the DC operation, an 1 s integration time would correspond to $10^8$ HV pulses in this case. Thus, the gas mixture is exposed to $10^5$ more HV pulses in the DC case, resulting in a much more intense luminous effluent. Interestingly, a particularly intriguing feature is revealed under ns-pulsed operation where the base of the effluent is surrounded by a "floating" conical luminous expansion formed just above the hole's exit. This feature could be evidence of additional electrically-driven effluent dynamics,[16] and is expected to affect the 2D profile of N–atoms density in the low pressure chamber (studied via ns-TALIF).

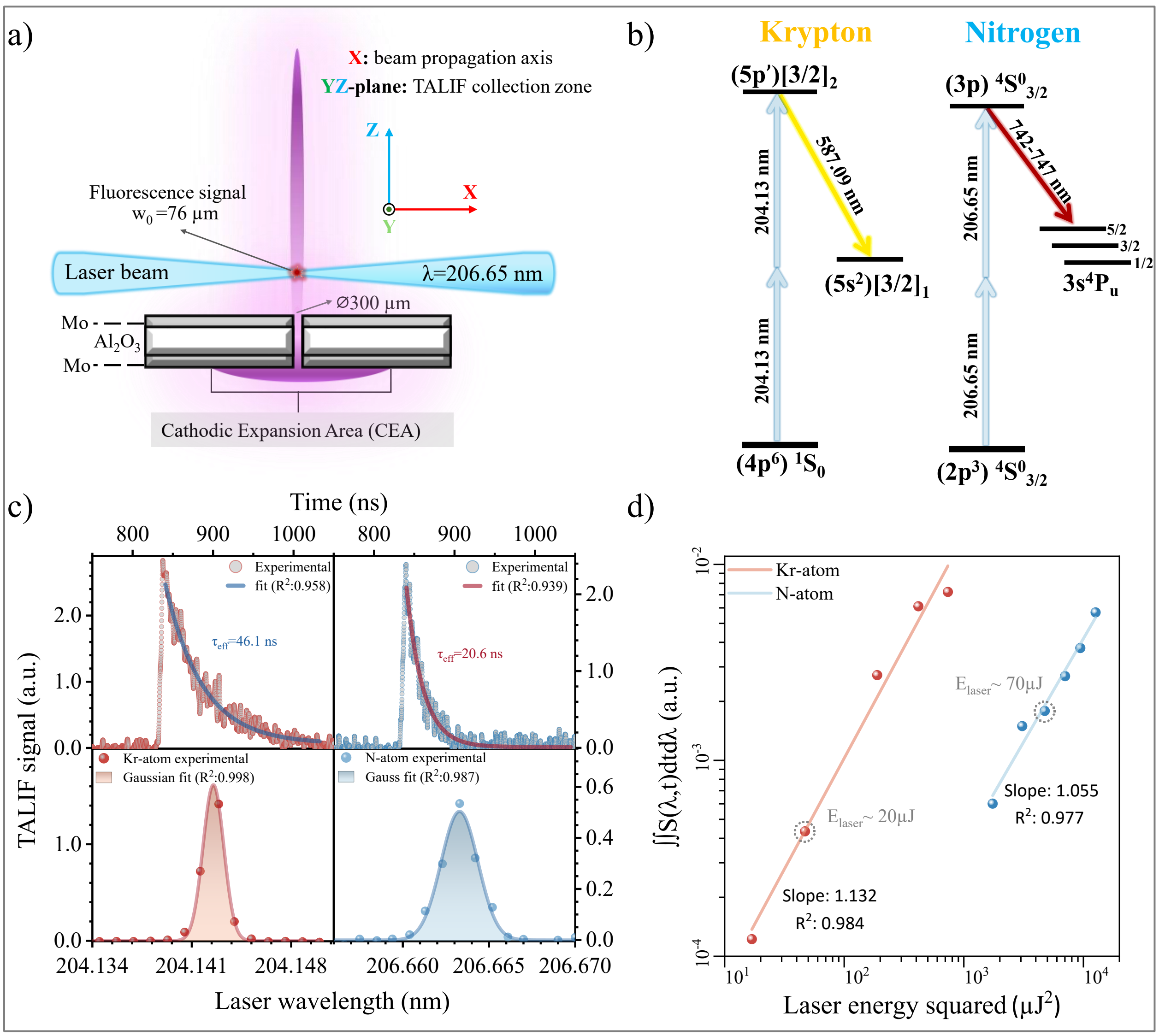

**FIG. 2.** (a) MHCD structure and laser beam's properties used for probing N–atoms. (b) TALIF schemes for Kr– and N–atoms. (c) Temporal TALIF signals of both species at resonance (top) and corresponding absorption line profiles (bottom). (d) Intensities of temporally- and spectrally-integrated TALIF signals versus the squared laser energy.

While a comprehensive description of the ns-TALIF setup is available elsewhere,[6,20] some essential details are provided here. Fig. 2(a) presents a simplified illustration of the MHCD structure along with the spatial laser beam profile used to map the N-atom density ($n_N$) in the low-pressure chamber. The absolute density is determined via a dedicated calibration procedure based on similar measurements in the same reactor filled with Krypton (Kr) gas, following the methodology established in.[6,21]

TALIF measurements are performed using a tunable dye laser (Sirah Lasertechnik Cobra-Stretch) pumped by a Q-switched Nd:YAG laser (Spectra Physics Quanta-Ray Lab-Series 170–10) operating at a repetition rate of 10 Hz. The system provides 7 ns UV laser pulses and a beam waist of 76 μm (**Fig. 2(a)**), measured as shown in.[7] To probe N–atoms, the laser wavelength is scanned over a 0.02 nm range centered at 206.65 nm (theoretical value[22]) and the fluorescence signal from the laser-excited state is collected in the spectral range 742–747 nm using a spectrometer (Model SP-2155, Princeton Instruments) coupled to a gated photomultiplier tube (PMT—Hamamatsu, H11526-20-NF, 2 ns temporal resolution). The associated TALIF scheme is illustrated in **Fig. 2(b)** along with that used for the calibrating gas (Kr). The fluorescence signal is recorded at multiple locations along the y and z axes in the low-pressure chamber [**Fig. 1(a)**]. The absorption line profiles for each species [**Fig. 2(c)**, bottom] are constructed by capturing temporal TALIF signals at various laser wavelengths around resonance [**Fig. 2(c)**, top]. These refer to selected laser pulse energies ($E_L$) which are within the so-called quadratic regime where TALIF measurements are valid [$E_L$=20 μJ for Kr–atoms and $E_L$=70 μJ for N–atoms; **Fig. 2(d)**]. By fitting the decaying phases of temporal TALIF signals with an exponential decay function [see, indicatively, **Fig. 2(c)**], their effective lifetimes are extracted and used for determining N–atom absolute densities.[6,20]

The electrical characteristics of an ns-pulsed MHCD, along with the synchronization between the HV and laser pulses, are shown in **Figure 3**. Representative waveforms of the incident ($V_{inc}$), reflected ($V_{refl}$), and net discharge voltage ($V_{dis}$=$V_{inc}$+$V_{refl}$) along with corresponding current signals during a single HV pulse are shown in **Fig. 3(a)**. Taking into account the peak values of $V_{inc}$ and $V_{refl}$ (reached around t=4.5 ns), the reflection coefficient ($R_c$) is estimated via $R_c=V_{refl}/V_{inc}\approx 0.9\ \text{kV}/1.9\ \text{kV}=0.47$, which helps in approximating the discharge impedance ($Z_{dis}$) at the same instant via $Z_{dis}=Z_0\frac{1+Rc}{1-Rc}\approx 50\frac{1+0.47}{1-0.47}\ \Omega\approx 140\ \Omega$ ($Z_0$=50 Ω is a typical coaxial cable impedance; see supplementary material **S1** for details).

Based on the calculated $Z_{dis}$, the electron density ($N_e$) can be estimated as follows:[23]

$$N_e = \frac{m_e v_m d}{Z_{dis} e^2 A} \qquad (1)$$

where $m_e$ is the electron mass, $v_m$ the elastic collision frequency between electrons and heavy particles, $d$ the height of the MHCD hole, $A$ the hole's cross-section, and $e$ the elementary charge. Using the above $Z_{dis}$ values, **eq. 1** provides an electron density of the order of a few $10^{14}$ cm$^{-3}$ which agrees qualitatively with previously-reported values on DC MHCDs.[5,10,11]

In **Fig. 3(a)**, $V_{dis}$ peaks at 3.0 kV within 5 ns, inducing a primary discharge current peak of nearly 20 A followed by lower-amplitude oscillations, typical of ns-pulsed discharges.[18,19] By recording electrical signals at various HV amplitudes, the temporal profiling of the corresponding energies ($E_{pulse}=\int_0^T V_{dis} I_{dis} dt$, $T$ being the HV pulse period) are obtained [**Fig. 3(b)**]. Here, the final value of $E_{pulse}$ reached after several reflections of the HV pulse [t=700 ns in **Fig. 3(b)**], is taken as the final deposited energy ($E_F$). **Fig. 3(b)** shows indicative $E_{pulse}$ profiles versus the HV amplitude (*f*=5 kHz). $E_{pulse}$ scales monotonically with $V_{inc}$, and $E_F$ increases from approximately 10 μJ at 2.0 kV to 95 μJ at 2.8 kV. This demonstrates a strong effect of the HV amplitude, since a mere 1.4–fold increase in $V_{inc}$ yields an approximate 9.5–fold increase in $E_F$.

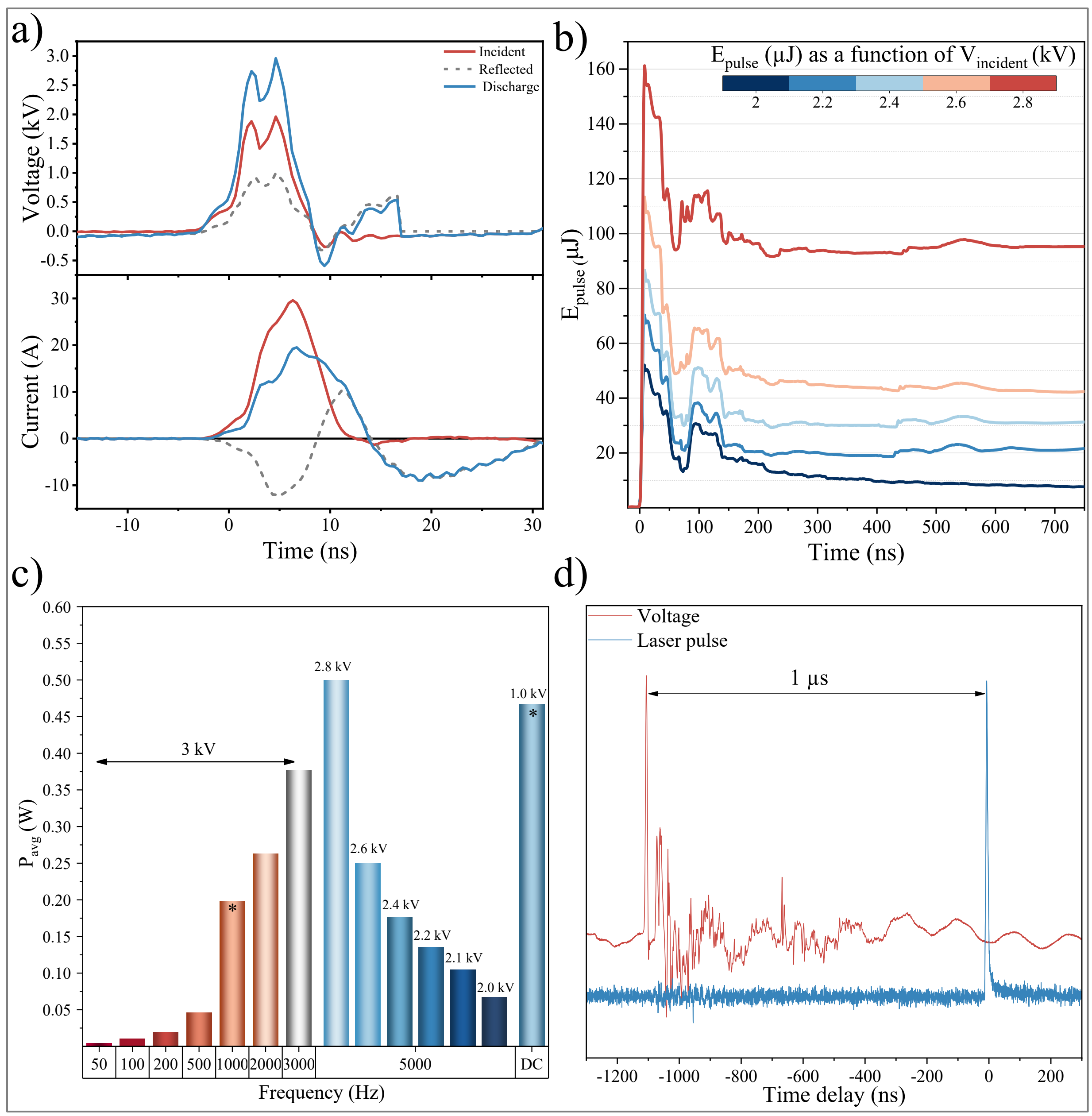

**FIG. 3.** Electrical features of the ns-pulsed MHCD. (a) Incident, reflected and net discharge voltage pulses, and corresponding induced currents. (b) Deposited pulse energy as a function of voltage amplitude at $f$=5 kHz. (c) Evolution of the average discharge power ($P_{avg}$) versus frequency and voltage amplitude under ns-pulsed operation compared to DC operation (bar shown on the far right). Asterisks denote conditions for which ns-TALIF is performed. (d) Time delay between the HV and laser pulses used for time-resolved TALIF measurements (see **Fig. S2.1** in the supplementary material). Operating conditions: 50%Ar–50%$N_2$, $P_{High}$=47 mbar, $P_{Low}$=13 mbar.

The limits of the electrical parameters of ns-pulsed MHCD for reliable and stable operation are detailed in **Fig. 3(c)**. This figure depicts the average consumed discharge power ($P_{avg}=\frac{1}{T}E_{pulse}$) over a range of $V_{inc}$ (2.0–3.0 kV) and $f$ (0.05–5 kHz). The power corresponding to the most stable condition under DC operation ($V_{inc}$=1 kV) is also shown for comparison. The consumed $P_{avg}$ in DC mode (0.47 W) is noticeably higher than most of the operating conditions of the ns-pulsed MHCD. In the ns-pulsed case, when $V_{inc}$ is held at 3.0 kV and $f$<3 kHz, $P_{avg}$ exhibits a sharp increase with frequency from 0.004 W at 50 Hz to 0.38 W at 3 kHz. This corresponds to a 95–fold rise on $P_{avg}$ for a 60–fold increase in frequency. Further increment in $f$ is restricted for $V_{inc}$=3.0 kV as it triggers undesirable instabilities and electromagnetic noise that compromise the MHCD operation. Nevertheless, safe operation can be extended to $f$=5 kHz by marginally reducing $V_{inc}$ to 2.8 kV, where the corresponding $P_{avg}$ reaches approximately 0.5 W.

A subsequent reduction in $V_{inc}$ (at $f$=5 kHz) leads to a steady decline in $P_{avg}$, which drops from 0.25 W ($V_{inc}$=2.6 kV) to 0.07 W ($V_{inc}$=2.0 kV). Throughout the parameter space tested, an overall peak $P_{avg}$ of 0.50 W is recorded at $f$=5 kHz and $V_{inc}$=2.8 kV, which is the only condition with higher $P_{avg}$ than the DC case. The globally smaller average power highlights the ns-pulsed MHCD as a highly cost- and energy-efficient technology for potential applications.

For the ns-pulsed MHCD, the temporal synchronization between the HV (<200 ps jitter) and laser (<500 ps jitter) pulses used for ns-TALIF is shown in **Fig. 3(d)**. A time delay of 1 μs is set between the two signals (see **Fig. S2.1** in the supplementary material for justification). For DC excitation no synchronization is needed.

The optical properties of DC and ns-pulsed MHCDs are investigated in different Ar/$N_2$ mixtures. The emitted light mainly originates in various Ar I lines and $N_2$ bands, while faint emissions from excited atomic nitrogen are also seen (see **Fig. S2.2** in the supplementary material). Furthermore, in DC-driven MHCDs, previous studies showed that the discharge morphology on the cathode is affected by the gas mixture ($N_2$, Ar, or Ar/$N_2$), pressure, and discharge current ($I_{dis}$). Notably, a quasi-symmetric expansion has been obtained in pure Ar which becomes asymmetric with increasing $N_2$ content in Ar/$N_2$,[7] in agreement with our results. Our studies further reveal that under ns-pulsed operation the cathodic expansion is symmetric and remarkably stable under all operating conditions which is an advantage for enhancing MHCD stability and longevity. Conventional photographs for different gas mixtures, voltages and pressures are provided in the supplementary material (**Fig. S2.3**).

**Figure 4** shows representative results in pure Ar when imposing an equal pressure of 30 mbar in both chambers. The discharge initiates within the micro-hole and expands across the cathode surface, closely resembling the normal glow regime of conventional DC-driven MHCDs.[10,11,13,24] To quantify this behavior, the cathodic expansion area (CEA) covered by the discharge is calculated for both HV waveforms. **Fig. 4(a)** shows conventional time-integrated photographs of the CEA under ns-pulsed excitation as a function of the $f$ for 3.5 and 2.2 kV. At 3.5 kV, increasing $f$ from 20 Hz to 1 kHz promotes a symmetric, circular expansion of the luminous glow. This expansion grows significantly at higher frequencies ($f$=5–10 kHz), even though $V_{inc}$ must be lowered to 2.2 kV to prevent device malfunction. By analyzing these photos, radial intensity profiles are extracted and fitted using a double-Gaussian convolution function, as representatively shown in **Fig. 4(b)**. The raw spatial profile (grey) is satisfactorily deconvoluted ($R^2$=0.998) into a narrow, highly intense core (orange), representing the microplasma confined within the MHCD hole, and a broad component (blue) capturing the diffuse luminous disk expanding across the cathode surface. For the illustrated case, this yields a characteristic expansion width r=1.92 mm which is taken as the radius of the circular CEA. Similar profiles are obtained under DC operation, though they exhibit slight asymmetricity at higher discharge currents [see, e.g., **Fig. 4(c)**], in agreement with previous studies.[7,11] Ultimately, the calculated CEA allows for the approximation of the average power density under each HV excitation by utilizing the corresponding calculated $P_{avg}$. Notably, while $P_{avg}$ is distributed between the internal discharge in the micro-hole and the extended discharge on the cathode, the volume of the MHCD hole is not considered for the CEA calculation in this study due to its 3D nature relative to the planar 2D cathode geometry.

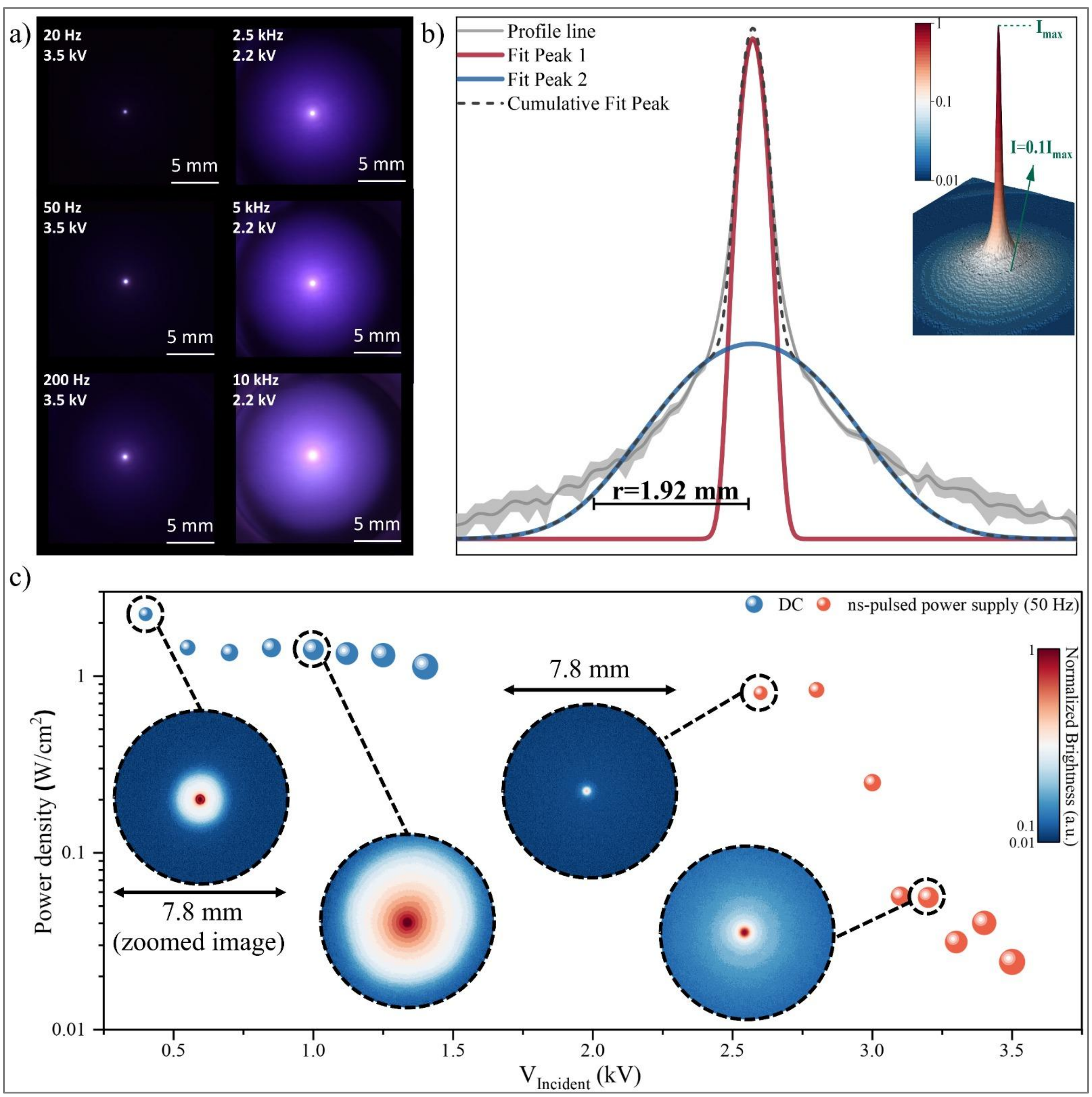

**FIG. 4.** Indicative variation of the CEA (in cm$^2$) at 100%Ar and 30 mbar pressure in both chambers. (a) Evolution of the CEA with frequency for two different $V_{inc}$ under ns-pulsed excitation. (b) CEA calculation method. (c) Power density (=$P_{avg}$/CEA) as a function of $V_{inc}$ under DC and ns-pulsed (f=50 Hz) operations.

Fig. 4(c) illustrates the calculated power density as function of $V_{inc}$ for both excitations. The false-color insets highlight the normalized brightness distribution across the cathode surface. For the DC excitation (blue spheres), the power density remains relatively stable above 1 W/cm$^2$ with a diffuse core which expands with increasing $V_{inc}$. Conversely, the ns-pulsed HV (orange spheres) reveals a strong, monotonic inverse relationship between the power density and the voltage. At lower $V_{inc}$ amplitudes (2.6 kV), the discharge is highly localized within the micro-hole. As $V_{inc}$ increases up to 3.5 kV, the cumulative profile diffuses noticeably outwardly, driving a sharp order-of-magnitude decrease in localised power density to below 0.05 $W/cm^2$. Thus, the ns-pulsed MHCD can reach significantly reduced power density compared to the DC mode, particularly at higher $V_{inc}$. This result reveals better energy efficiency of the ns-pulsed MHCD. In particular, by utilizing low pulse repetition frequencies ($f$=50 Hz in this example), this mode minimizes thermal stress on the cathode while efficiently channeling electrical energy into electron-driven reactions. This is also true for frequencies as high as 5 kHz as long as $V_{inc}\leq 2.6$ kV (see **Fig. 3(c)**). Consequently, device longevity is substantially improved while maintaining enhanced reactivity. In fact, conventional DC operation leads to irreversible defects within a maximum operating time of a few days at best (non-continuous full-day operation), while ns-pulsed MHCD enables smooth full-day operation for at least 10 days (corresponding to the duration of the experiments; this can be even higher) without any critical defect.

The spatial distribution of the absolute N–atom density ($n_N$) measured via ns–TALIF is shown in **Figure 5** for both power supplies. Variation along the jet centerline [y=0 mm, **Fig. 5(a)**] reveals that ns-pulsed excitation sustains a consistently higher N–atom density that shows only a minor decrease downstream, dropping from approximately $4\times10^{14}$ $cm^{-3}$ (z=5 mm) to only $2\times10^{14}$ $cm^{-3}$ (z=30 mm). In contrast, the DC-driven profile exhibits greater fluctuation. Although the density peaks near $4\times10^{14}$ $cm^{-3}$ close to the hole exit (z=0 mm), it drops sharply by nearly an order of magnitude to approximately $5\times10^{13}$ $cm^{-3}$ at z=30 mm. This distinct behavior indicates that ns-pulsed excitation achieves superior downstream transport of N–atoms, pointing to enhanced initial dissociation rates coupled with lower recombination losses. The spatial divergence is further clarified by the radial profiles taken at an intermediate downstream position [z=15 mm, **Fig. 5(b)**]. Under ns-pulsed operation, $n_N$ forms a broad, quasi-uniform plateau across the central core ($-5\leq y\leq +4$ mm) at an elevated atomic density of (4–6)$\times10^{14}$ $cm^{-3}$, with a slight upward trend near the outer shear layer boundary ($y\geq +5$ mm; $n_N\approx 8\times10^{14}$). Conversely, the DC profile displays a parabolic (or bell-shaped) distribution that peaks within a smaller core region ($-2\leq y\leq +2$ mm) at a significantly lower atomic density ($n_N\approx 10^{14}$ $cm^{-3}$) and falls off drastically toward the periphery ($n_N<5\times10^{13}$ $cm^{-3}$).

The complete 2D normalized density mapping over the investigated region [**Fig. 5(c)**] provides a preliminary overview of N–atoms topography, superimposed with various discrete experimental data points. The 2D profiles are obtained using a linear interpolation ($10^3$ points) between adjacent discrete points. Even if the data are normalized to an absolute density value of $5.72\times10^{14}$ $cm^{-3}$ (to facilitate comparison), which corresponds to half of the maximum overall density measured in this work (i.e., $n_N=1.14\times10^{15}$ $cm^{-3}$ at y=17 mm and z=15 mm under ns-pulsed operation), a remarkable difference is revealed between the two operating regimes. The DC density zone (left) remains mostly localized with relatively lower values restricted to a central core axis. In contrast, the ns-pulsed density zone (right) exhibits a vast expansion where substantial N–atoms accumulation (>60% of maximal density) extends into the peripheral shear layers (y>12 mm) and downstream zones (e.g., z=15 mm). This view underscores the ability of high-peak-power ns-pulsed operation to fundamentally alter the spatial structure of molecular dissociation compared to conventional steady-state DC operation.

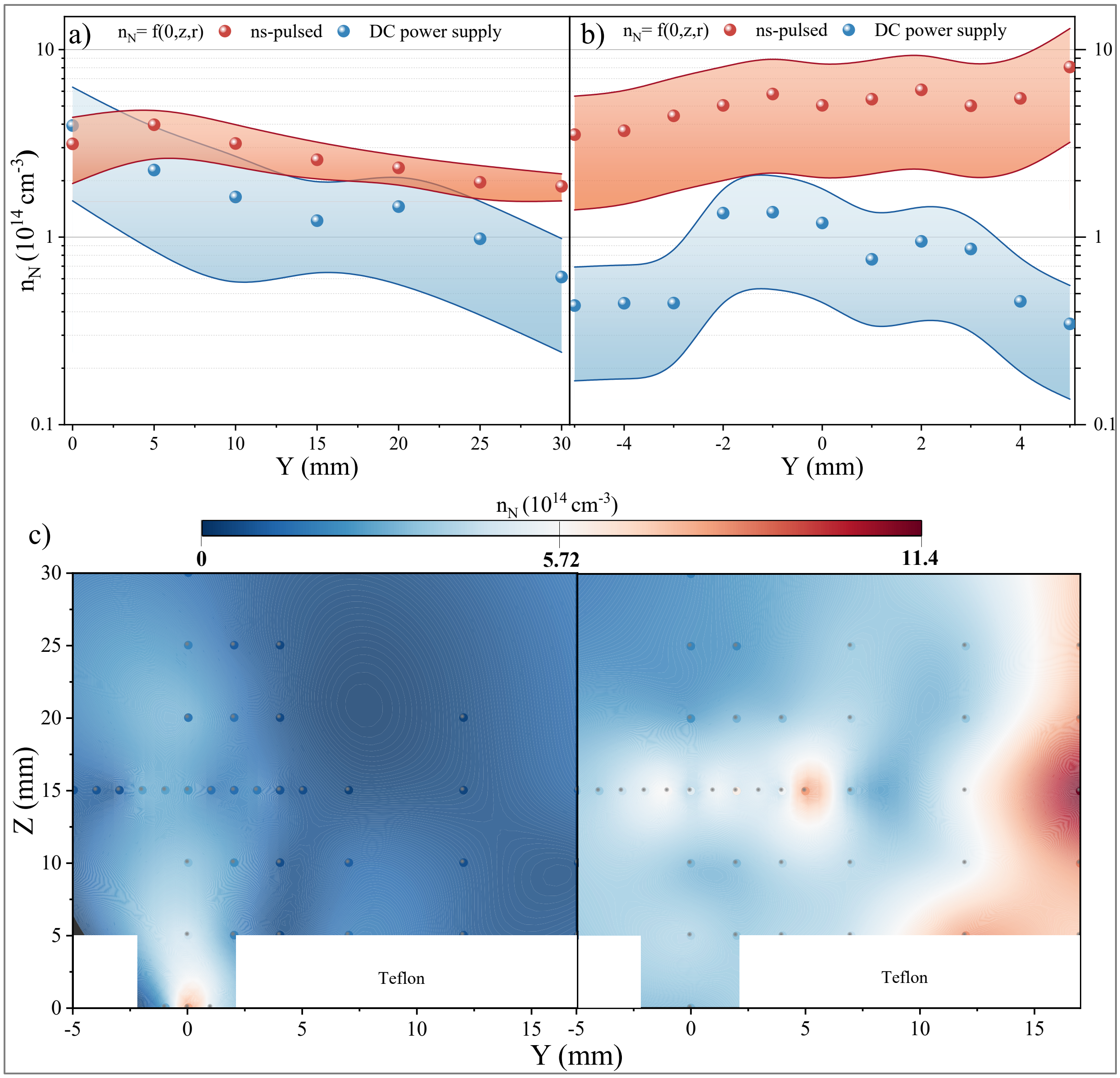

**FIG. 5.** (a) ns-TALIF measurements of absolute N–atom density ($n_N$) along the jet centerline (y=0 mm) and (b) along the radial direction (z=15 mm), for both power supplies. (c) 2D partial mappings of the normalized N–atom density in the low-pressure chamber for DC (left) and ns-pulsed (right) excitations. The spheres represent the experimental points used to generate the interpolated 2D spatial distributions.

A striking difference emerges when correlating the time-integrated photos of the luminous effluent [**Fig. 1(b)**] with the corresponding absolute density mappings [**Fig. 5(c)**] under each voltage regime. While DC discharge exhibits a relatively longer, bright and focused plume, its radial profile remains tightly confined to a narrow core with increasingly lower lateral intensity. The shape of the luminous pattern is dominated by the gas flow dynamics forming a strong downstream jet which transports excited species from the active discharge (in the hole) to the low-pressure chamber. Thus, the effluent represents the discharge afterglow which comprises majoritaly excited and ground-state neutrals.[12] This is confirmed by performing ns-TALIF mappings in Kr gas showing similar morphology (see **Fig. S2.4** in supplementary material). Conversely, the ns-pulsed discharge is created within an extremely short time duration (<10 ns). Thus, the plasma remains "OFF" for the vast majority of the time interval between consecutive pulses (here, 1 ms at $f$=1 kHz). This leads to a noticeably shorter jet, manifesting as a highly-localized bright core surrounded by a large conical luminous expansion just after the hole exit [**Fig. 1(b)**]. However, the strong induced electric field by the ns-pulsed HV channels energy with high efficiency into direct molecular dissociation rather than continuous optical emission.[25,26] This creates a dense, long-lived N–atom population that accumulates and quickly expands into a massive downstream distribution, which matches qualitatively the broad, conical expansion appearing in the ns-pulsed photograph [**Fig. 1(b)**; right]. This behavior could be related to the fact that fast high-energy deposition in the case of ns-pulsed operation can generate strong density waves at the onset of each HV pulse. This effect is intensified by the micrometric hole diameter of the MHCD, which acts as a micro-nozzle restricting gas expansion and choking the transition flow.[12] Consequently, these transient waves travel downstream into the low-pressure chamber, altering the gas profiles and contributing to the broad, volumetric expansion of N–atoms observed in the ns–TALIF mappings.

In conclusion, this work demonstrates the successful implementation of a true ns-pulsed high-voltage excitation to drive an Ar/$N_2$ MHCD, offering a high-performance alternative to conventional DC systems. Through optical emission, electrical and space-resolved ns-TALIF diagnostics, we established that transient nanosecond pulsing fundamentally decouples time-integrated optical emission from chemical dissociation efficiency. While consuming significantly lower average power (≤0.50 W) and minimizing destructive thermal stresses, the elevated reduced electric field enhances dissociation pathways. This yields up to a six-fold enhancement in absolute ground-state N–atom density (maximum overall value measured: $1.14\times10^{15}$ $cm^{-3}$), driving a massive volumetric expansion downstream that is likely aided by transient micro-nozzle pressure waves. By overcoming the critical bottleneck of the triple bond of $N_2$ while simultaneously mitigating instabilities and source degradation, ns-pulsed MHCDs drastically improve device longevity (>10 days of continuous operation without any obvious defect) while functioning as high-yield N–atom sources. These findings unlock a promising path forward for the efficient and controlled generation of reactive N–atoms required in the synthesis of strategic nitride materials such as h-BN.

See the supplementary material for additional details on the determination of the discharge impedance ($Z_{dis}$) (**S1**), time-resolved TALIF densities of N–atoms (**S2; Fig. S2.1**), optical emission spectra (**S2; Fig. S2.2**), conventional photographs of the discharge expansion on the cathode (**S2; Fig. S2.3**), and 2D spatial mapping of Kr–atoms via ns-TALIF (**S2; Fig. S2.4**).

This work was supported by the French Research National Agency (ANR) via the project SPECTRON (No. ANR-23-CE51-0004-01) and LabEx SEAM (Nos. ANR-10-LABX-0096 and ANR-18-IDEX-0001). The authors would like to thank Ludovic William and Nicolas Fagnon for technical drawing and support.

## AUTHOR DECLARATIONS

### Conflict of Interest

The authors have no conflicts of interest to disclose.

## DATA AVAILABILITY

The data that support the findings of this study are available from the corresponding authors upon reasonable request.

# SUPPLEMENTARY MATERIAL

## Proof of operation of a micro-hollow cathode discharge driven by a true ns–pulsed high voltage: emission properties and atomic nitrogen production

### S1. Determination of the discharge impedance ($Z_{dis}$) based on transmission line theory

In this work, $Z_{dis}$ was determined from the incident and reflected voltage pulses measured during ns-pulsed operation [**Fig. 2(a)**]. Because the pulse duration was comparable to the electromagnetic transit time of the system, the MHCD reactor cannot be treated as a lumped circuit element and must instead be analyzed as a transmission-line load.

The pulse generator was connected to the MHCD reactor through a coaxial cable of characteristic impedance $Z_0$=50 Ω. When a voltage pulse propagating along the transmission line reaches the reactor, a fraction of the pulse is transmitted into the reactor while the remaining fraction is reflected due to the impedance mismatch between the reactor impedance ($Z_{dis}$) and $Z_0$. The total discharge voltage ($V_{dis}$) is given by the superposition of the incident ($V_{applied}$) and reflected ($V_{refl}$) waves, as follows:

$$V_{dis}(t) = V_{applied}(t) + V_{refl}(t) \tag{S1}$$

Similarly, the discharge current is given by:

$$I_{dis}(t) = [V_{applied}(t) - V_{refl}(t)]/Z_0 \tag{S2}$$

Adding and subtracting eqs. (S1) and (s2) we obtain:

$$V_{dis} + Z_0 I = (V_{applied} + V_{refl}) + (V_{applied} - V_{refl}) = 2V_{applied} \tag{S3}$$

$$V_{dis} - Z_0 I = (V_{applied} + V_{refl}) - (V_{applied} - V_{refl}) = 2V_{refl} \tag{S4}$$

Thus,

$$V_{applied}(t) = \frac{V_{dis}(t) + Z_0 I(t)}{2} \tag{S5}$$

$$V_{refl}(t) = \frac{V_{dis}(t) - Z_0 I(t)}{2} \tag{S6}$$

The voltage reflection coefficient is given by $R_c$=$V_{refl}(t)/V_{applied}(t)$. By substituting eqs. (S5) and (S6) we obtain:

$$R_c = \frac{V_{dis}(t) - Z_0 I(t)}{V_{dis}(t) + Z_0 I(t)} \quad \text{(S7)}$$

Finally, the discharge impedance is given by $Z_{dis} = V_{dis}/I_{dis}$. Substituting eqs. (S1) and (S2) we obtain:

$$Z_{dis} = Z_0 \frac{1 + V_{refl}/V_{applied}}{1 - V_{refl}/V_{applied}} = Z_0 \frac{1 + R_c}{1 - R_c} \quad \text{(S8)}$$

## S2. Optical emission spectra – Conventional photographs of the discharge expansion on the cathode – Additional TALIF data

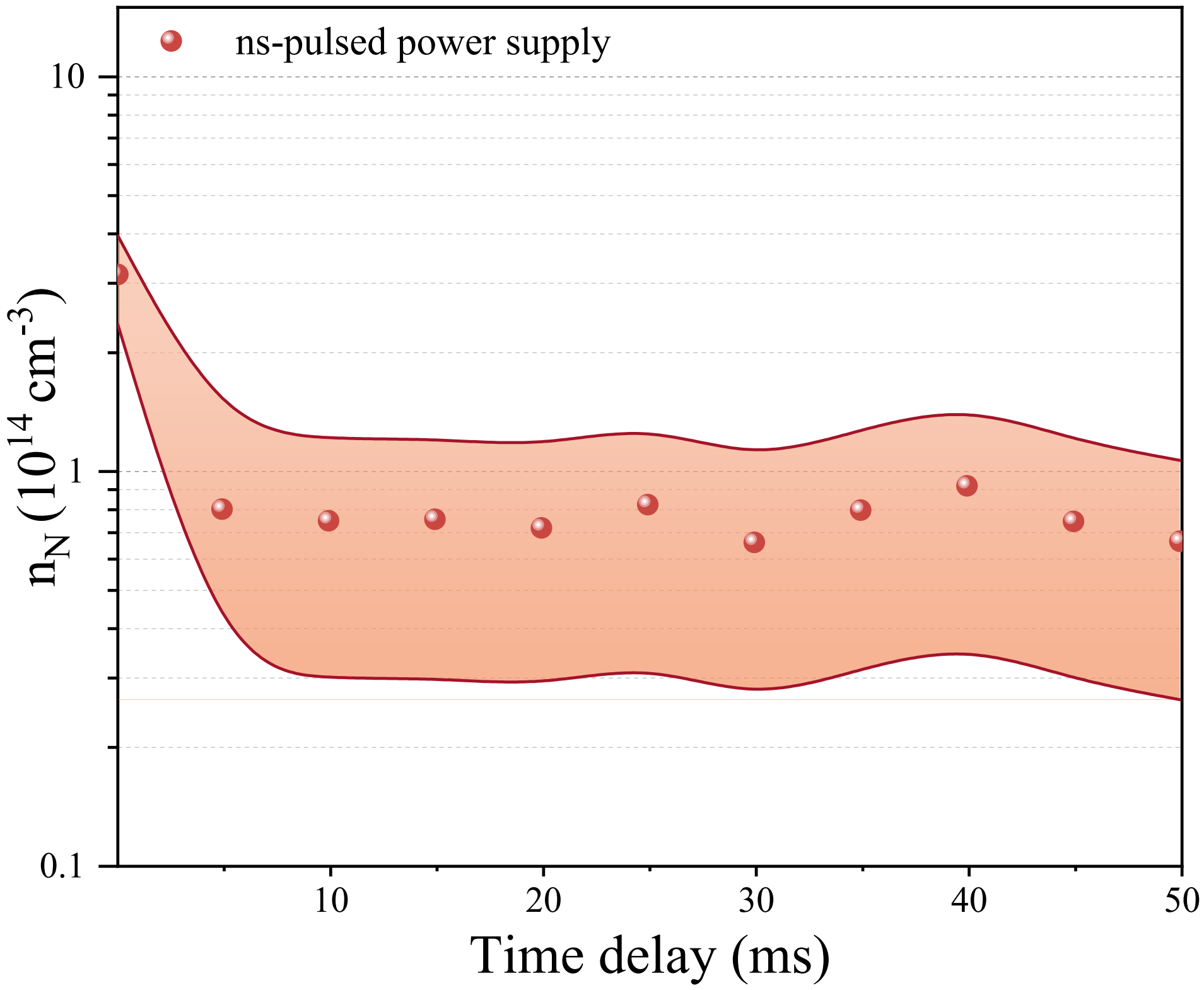

**FIG. S2.1** Time-resolved N–atom absolute density in the case of ns-pulsed MHCD. The error bars shown in light orange represented the systematic error of the method. For standard TALIF studies, a delay generator (DG645) is used to synchronize the laser with the HV power supply and the PMT gate as shown in **Fig. 3(d)**]. A time delay of 1 μs is set between the two signals. This ensures that N–atoms are probed after the active discharge phase (i.e., when the plasma is "OFF"), while effectively isolating the fluorescence signals from eventual strong discharge emissions (e.g., $N_2$(FPS)) occurring at the same wavelength as the TALIF signal. From this study, a strong TALIF signal is found at t=1 μs which decreases progressively at higher time instants. The laser serves as the master trigger for all timing events. The DG645 relays this signal to the PMT, which opens at the same frequency as the laser (10 Hz). To synchronize the laser with the pulsed HV signal (operating at a different frequency), the burst mode of the DG645 is used. This allows scanning 1 kHz burst pulses of DG645 (100 bursts) with the laser TTL (10 Hz). For time-resolved TALIF measurements of ground-state nitrogen atoms, we maintained the same HV pulse frequency but reduced the burst number to 50. By temporally shifting the voltage pulses relative to the laser TTL, we directly probed the nitrogen fluorescence signal after the HV pulse (i.e., when the discharge is "OFF").

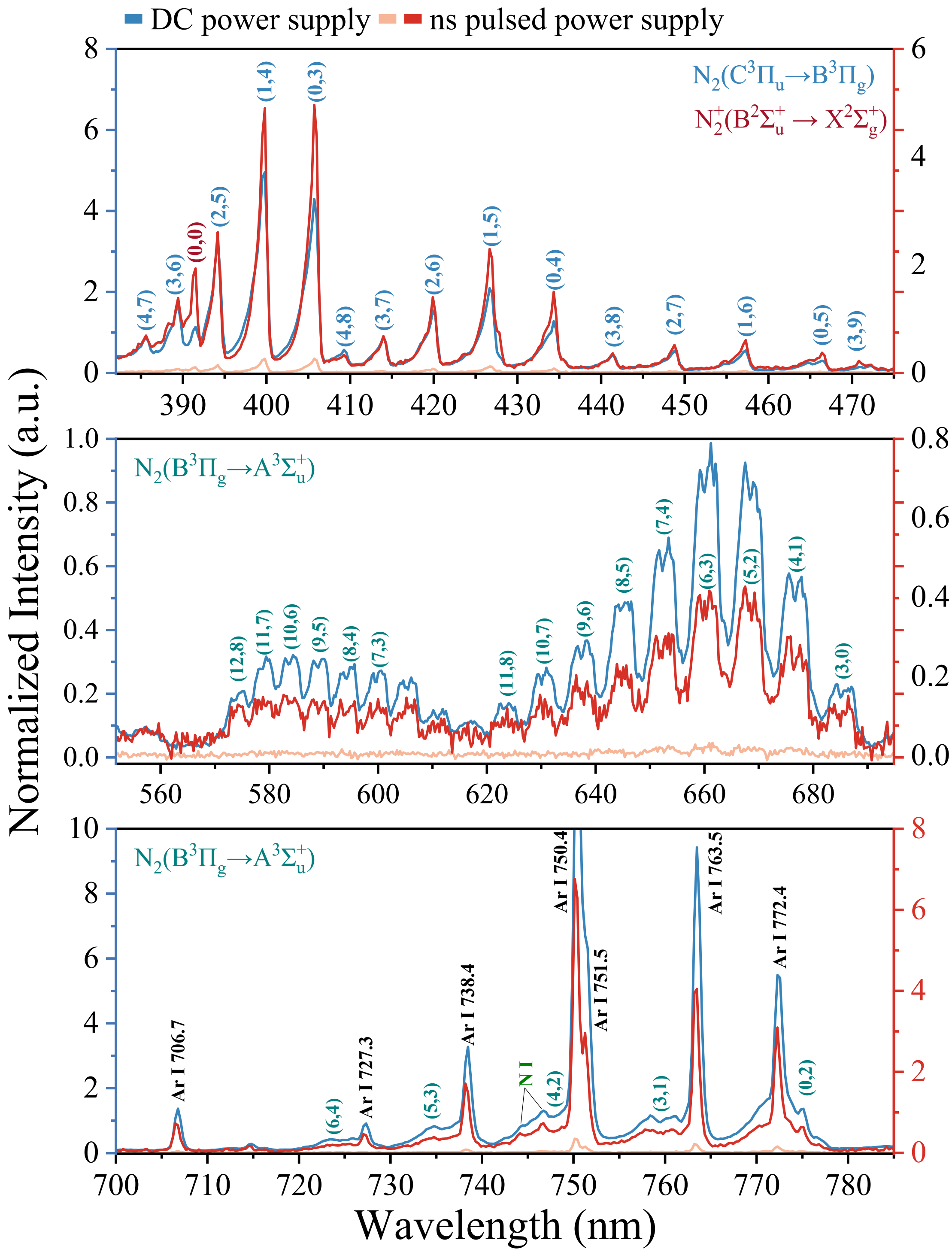

**FIG. S2.2** Superposition of the space-integrated (originating from the MHCD hole and the effluent) and time-integrated (500 ms integration time for the orange and blue spectra) optical emission spectra under ns-pulsed (orange) and DC (blue) operation. The red spectra correspond to 10 s integration time in the case of ns-pulsed operation to improve visibility. Two very weak N I lines (shown in green) are distinguished around 744 and 746 nm in both cases.

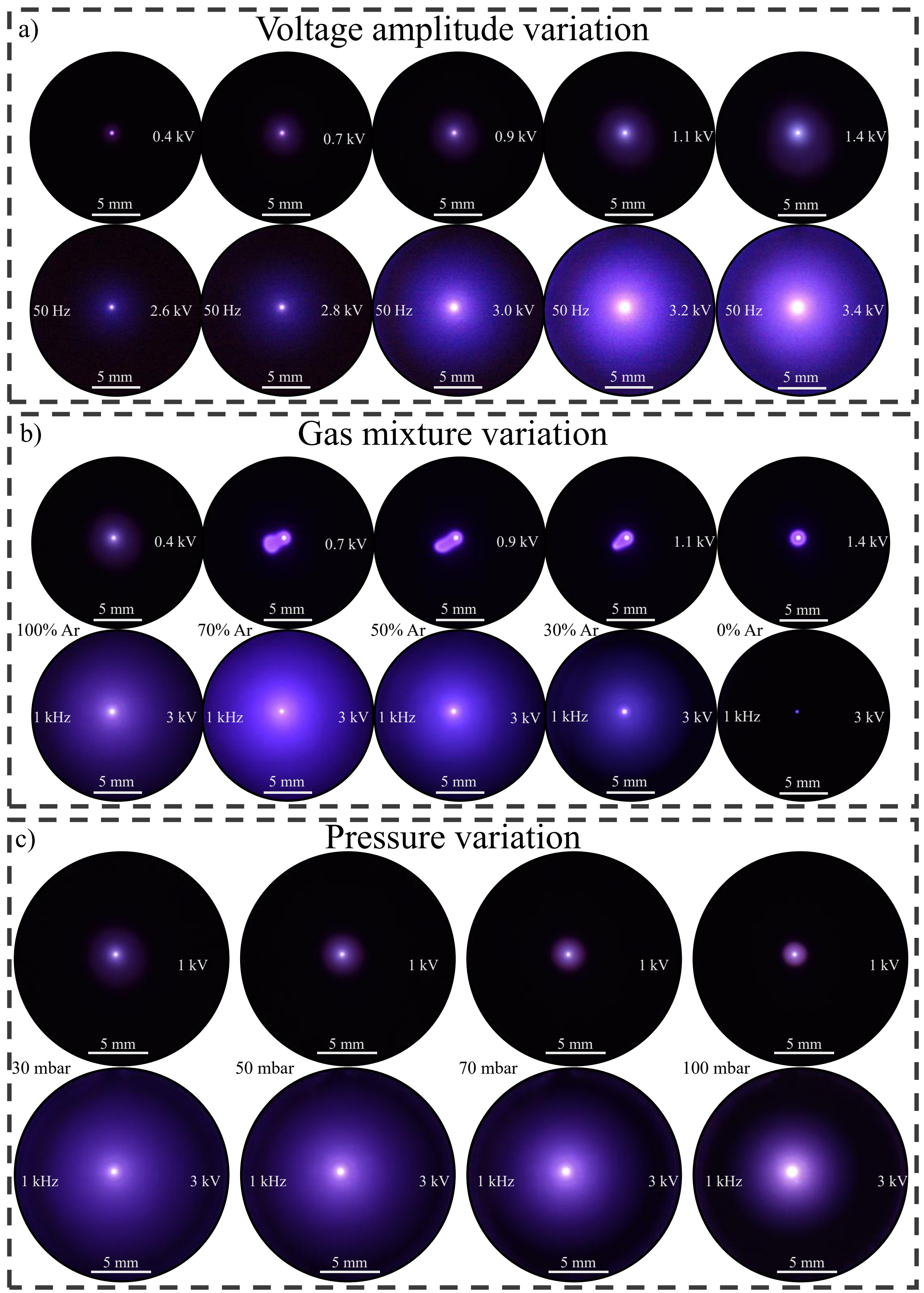

**FIG. S2.3** DSLR photos of the CEA under different incident voltage amplitudes ($P_{high}$=$P_{low}$=30 mbar) (a), Ar/$N_2$ gas mixtures ($P_{high}$=$P_{low}$=30 mbar) (b), and gas pressures (same $P_{high}$ and $P_{low}$) (c) for both power supplies. The top and bottom rows in (a), (b) and (c) refer to DC and ns-pulsed operation, respectively.

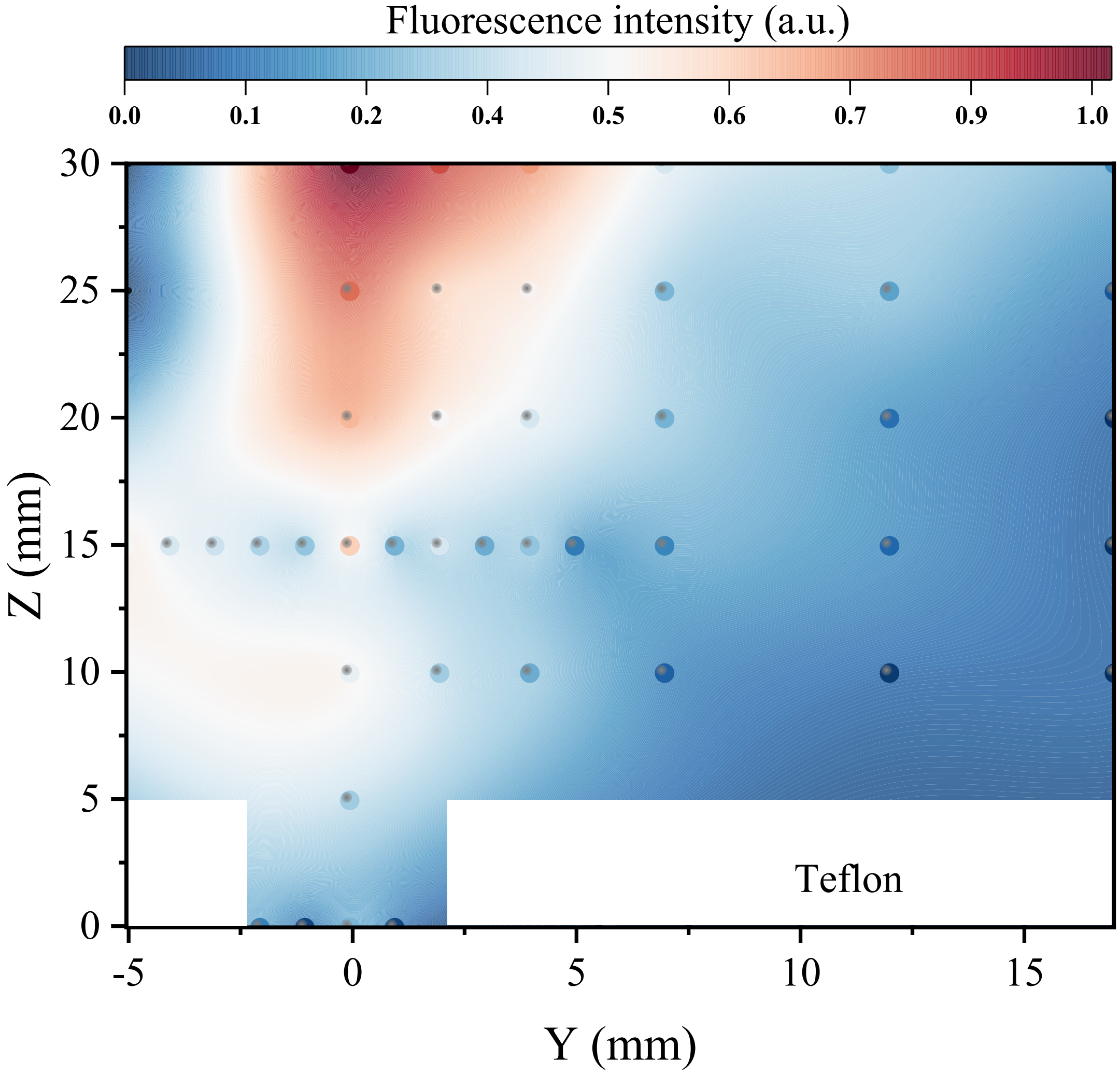

**FIG. S2.4** Kr–atom ns-TALIF mapping in the low-pressure chamber ($P_{high}$=10×$P_{low}$). These are used to visualize the effect of the gas flow on the transport of fluorescing atoms in the low-pressure chamber, which is expected to happen in N–atoms produced with the DC MHCD.